\documentclass[aps,pre,onecolumn,groupedaddress]{revtex4}
\usepackage{graphicx}
\usepackage{bm}
\usepackage{amsmath, amsthm, amssymb}
\usepackage{eurosym}
\usepackage{tabularx}

\begin{document}

\title{Synchronization and Anti-synchroniztion of Dynamically Coupled
Networks}
\author{M. Turalska$^{1}$}
\author{A. Svenkeson$^{2}$}
\author{B.J. West$^{1,3}$}
\affiliation{
$^{1}$ Physics Department, Duke University, Durham, NC 27709, USA\\
$^{2}$ CISD, Army Research Laboratory, Adelphi, MD\\
$^{3}$ Information Science Directorate, US Army Research Office, Research
Triangle Park, NC 27708, USA
}
\date{\today }

\begin{abstract}
We consider the coupling between two networks, each having N nodes whose
individual dynamics is modeled by a two-state master equation. The
intra-network interactions are all to all, whereas the inter-network
interactions involve only a small percentage of the total number of \ nodes.
We demonstrate that the dynamics of the mean field for a single network has
an equivalent description in terms of a Langevin equation for a particle in
a double-well potential. The coupling of two networks or equivalent coupling
of two Langevin equations demonstrates synchronization or
antisynchronization between two systems, depending on the sign of the
interaction. The anti-synchronized behavior is explained in terms of the
potential function and the inter-network interaction. The relative entropy
is used to establish that the conditions for maximum information transfer
between the networks are consistent with the Principle of Complexity
Management and occurs when one system is near the critical state. The
limitations of the Langevin modeling of the network coupling are also
discussed.
\end{abstract}

\maketitle

\section*{Introduction}
The significance of synchronization for the understanding of complex systems became evident to the broad scientific community with the publication of Strogatz's remarkable book \textit{Sync} \cite{strogatz03}. Although the underlying mechanisms can be quite varied in different kinds of events synchronization is central in phenomena from the coordinated clapping of an audience, to the rhythmic firing of pacemaker cells in the heart, to the syncopated firing of neuronal groups. Tognoli and Kelso \cite{tognoli08} identify three types of synchrony involving relative phases between coupled regions of the brain: an inphase (zero-lag) synchronization; antiphase in which oscillatory elements have the same intrinsic frequency; and finally broken symmetry that is near inphase or near antiphase.

Phase synchronization is not only a fundamental mechanism giving rise to a collective action of large ensembles of units found in near proximity to each other, which otherwise would behave according to their individual rhythms, not being able to produce a large scale response.\cite{holst39}. An inphase pattern has been observed in brain activity using fMRI measurements within a cohort of patients by synchronizing brain activity across individuals watching the same movie \cite{nummenmaa12}. The inphase pattern of the activity in different regions of the brain among different members of the audience cohort reveals shared emotional states.

An antiphase pattern in the absence of a task has been explained by Li and Zhou \cite{li11} as a type of background organized by the spontaneous cortico-cortical communication dynamics. Their two module coupling calculations using both neuron cell and neuron mass models suggest that the antiphase dynamics observed in both is generic. The challenge is then to explain the anti-synchronization property independently of specific mechanisms.

In nature, many complex networks are modular, composed of subnetworks, with varying internal and external connectivity. Here we study the interaction of two modules (networks), both having the same structure and functionality. We demonstrate that cooperative dynamics of single units within each module leads to consensus. When coupled, dynamic consensus can act to counter the influence of the inter-network coupling, resulting in an anti-synchronization of the two networks or it can act to synchronize the cooperative impulse. The response of the two networks to one another depends on their internal states and on the symmetry of the interaction. The symmetry of the interactive response has been discussed in the context of coupled chaotic systems as an adaptive control scheme \cite{ahn09,zhang10}. Herein the symmetry of the interactive networks is the result of a control process that is dependent on the nature of the coupling of the networks. From this we establish consistency with the Principle of Complexity Management (PCM) \cite{west08,aquino11}, thereby suggesting the conditions allowing for maximum information transfer across complex networks composed of multiple modules.

In Section \ref{DMM} the decision making model (DMM) is introduced in terms of the two-state master equation. A number of the DMM properties are reviewed, such as in the all-to-all (ATA) coupling of the elements in an infinite sized network the dynamics of the mean field variable are described analytically by a particle in a double-well potential and the magnitude of the potential barrier is a function of a cooperation parameter. For a finite sized network this description reduces to a Langevin equation and the strength of the fluctuations scales with the size of the network and is also a function of the cooperation parameter.

The coupling between two ATA DMM networks is considered in Section \ref{network response} and shown to be equivalent to two coupled Langevin equations. The strength of the fluctuations in the Langevin model are estimated from a corresponding DMM calculation thereby making the Langevin description empirical. The sign of the coupling terms determines the symmetry of the coupled network, that is, whether the dynamics of the coupled networks are synchronized or anti-synchronized. Moreover, the conditions under which the maximum information is shuttled back and forth between the modules are determined using relative entropy and it is determined that the information transfer is not dependent on the symmetry of the dynamics.

We suggest that in one possible application of the model the individual subnetworks can represent the cognitive behavior of individuals trying to make a decision between two alternatives. The potential minima in the Langevin model could then be associated with speaking and listening in a two person conversation. The distribution of time intervals spent with one person talking and the other listening is herein shown to be inverse power law consistent with prior research on turn-taking dynamics in which inverse power law was obtained \cite{pincus14}. Moreover, Abney \textit{et al}. \cite{abney14} have shown that the inter-event intervals ranging from phonemes to semantics in dyadic conversations have an inverse power law distribution with an index near -2 in keeping with the PCM for the maximum transfer of information between two complex networks.

In Section \ref{discussion} we draw some conclusions.

\section*{Methods \label{DMM}}

\subsection*{The Decision Making Model}

The Decision Making Model (DMM) implements the echo response hypothesis, which assumes that the dynamic properties of a network of identical individuals are determined by singular people imperfectly copying the behavior of one another \cite{west14}. Formally an isolated individual is modeled as an unit switching back and forth between two states, $+1$ and $-1$, with constant rate $g_{0}$ of making a transition at any time. Thus, the probability of finding an isolated unit $s(t)$ in one of two states, $\mathbf{p}(t)=(p_{+1},p_{-1})$ is described by a two-state master equation,

\begin{equation}
\frac{d\mathbf{p}(t)}{dt}=\mathbf{Gp}(t),  \label{indiv}
\end{equation}%
where $\mathbf{G}$ is a $2\times 2$ transition matrix with constant elements:

\begin{equation*}
\mathbf{G}=\left[
\begin{array}{cc}
-g_{0} & g_{0} \\
g_{0} & -g_{0}%
\end{array}%
\right] .
\end{equation*}

Here the adopted network structure is that of an all-to-all (ATA) network, in which each individual $s^{(i)}(t)$ interacts with all other individuals composing the network. The interactions modify the two-state master equation describing a single individual (Eq. \ref{indiv}) and lead to a master equation with time-dependent transition rates:

\begin{equation}
\frac{d\mathbf{p}^{\left( i\right) }(t)}{dt}=\mathbf{G}^{\left( i\right) }(t)%
\mathbf{p}^{\left( i\right) }(t),  \label{master}
\end{equation}%
where $\mathbf{G}^{\left( i\right) }(t)$ contains the transition rates of switching from state $+1$ to $-1$, $g_{+1\rightarrow -1}$, and the rates of switching from state $-1$ to $+1$, $g_{-1\rightarrow +1}$:
\begin{equation}
\mathbf{G}^{\left( i\right) }(t)=\left[
\begin{array}{ll}
-g_{_{+1\rightarrow -1}}^{\left( i\right) }(t) & g_{_{-1\rightarrow
+1}}^{\left( i\right) }(t) \\
g_{_{+1\rightarrow -1}}^{\left( i\right) }(t) & -g_{_{-1\rightarrow
+1}}^{\left( i\right) }(t)%
\end{array}%
\right] .  \label{coupling}
\end{equation}%
The probability of individual $i$ being in one of two states $(+1,-1)$, is $\mathbf{p}^{\left( i\right) }(t)=(p_{+1}^{\left( i\right) },p_{-1}^{\left(i\right) })$. Positioning $N$ such individuals at the nodes of a network yields a system of $N$ coupled two-state master equations \cite{turalska09,bianco08}. The transition rates for each of the $i$ individuals:%
\begin{eqnarray}
g_{_{+1\rightarrow -1}}^{\left( i\right) }(t) &=&g_{0}\exp \left[ -\frac{K}{N%
}\left( N_{+1}(t)-N_{-1}(t)\right) \right] ;  \notag \\
g_{_{-1\rightarrow +1}}^{\left( i\right) }(t) &=&g_{0}\exp \left[ \frac{K}{N}%
\left( N_{+1}(t)-N_{-1}(t)\right) \right]   \label{rates}
\end{eqnarray}%
depend on the the total number of individuals in the state $+1$ and $-1$, $N_{+1}(t)$ and $N_{-1}(t)$, respectively. Since every individual in the network stochastically chooses their state the total number of individuals within each state, $N_{+1}(t)$ and $N_{-1}(t)$ also fluctuates in time. As the number of the elements in the network increases and approaches $N=\infty$, the ratio $N_{\pm 1}(t)/N\rightarrow p_{\pm 1}\left( t\right) $, and the relative frequency becomes the probability that the network is in one state or the other. In this limit the transition rates become exponentially dependent on the state probabilities, resulting in a highly nonlinear master equation \cite{west14}. The parameter $K\ $denotes the strength of cooperation between elements of the network, being a measure of the extent individuals copy the behavior of one another.

\subsection*{Infinite ATA Network}

\begin{figure}[t]
\includegraphics{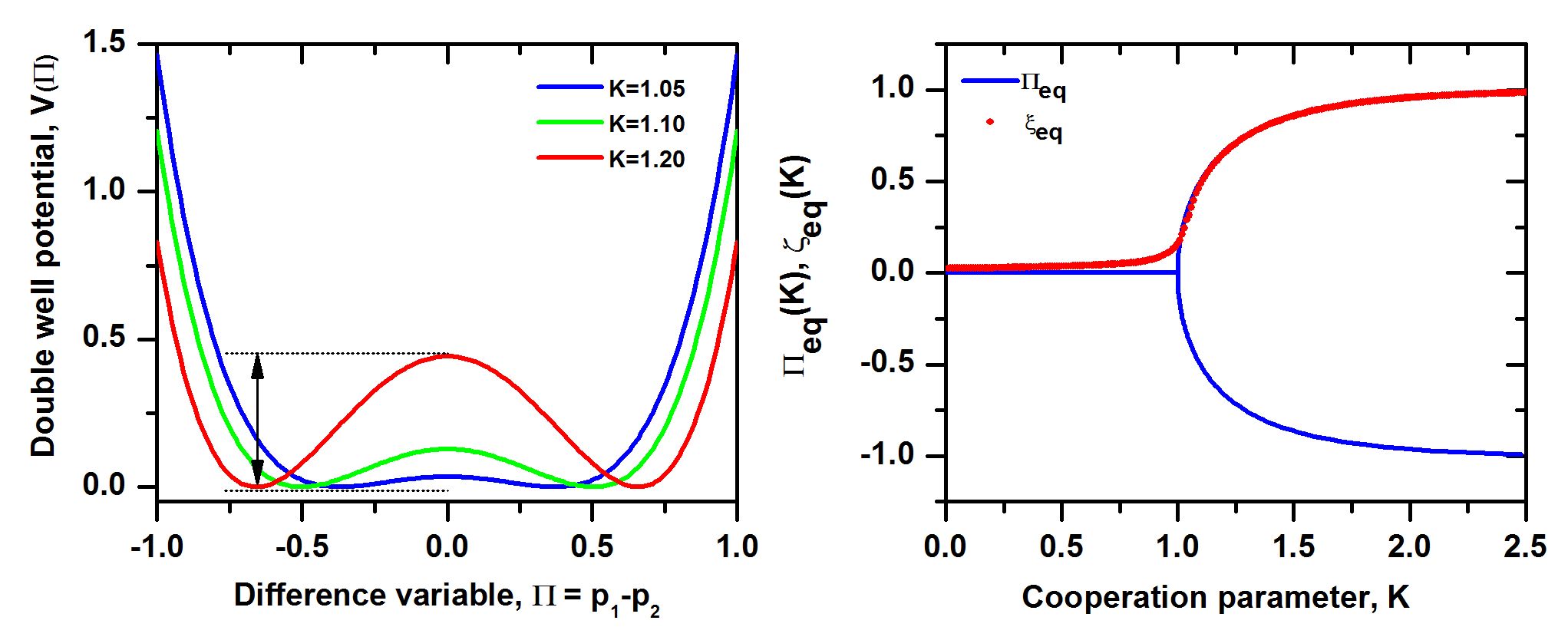}
\caption{{\bf Over-damped motion of a particle in the double-well potential $V(\Pi )$ is equivalent to the behavior of the mean field variable $\Pi $ (Eq. \protect\ref{global master}).}
\textit{(left)} The double well potential $V(\Pi )$ as a function of the difference variable $\Pi $ for three increasing values of the cooperation parameter $K$. An arrow marks the intensity of the barrier created between wells. \textit{(right)} The equilibrium value of the difference variable $\Pi _{eq}$, as a function of the cooperation parameter $K$. The thick line is the theoretical solution obtained by solving $d\Pi /dt=0$. The dots correspond to the numerical evaluation of $\protect\varsigma _{eq}$ for all-to-all network of $N=1000$ nodes. In both cases the unperturbed rate is $g_{0}=0.01$.}
\label{fig_well}
\end{figure}

In an ATA network of infinite size we may introduce the difference variable $\Pi ^{\left( i\right) }=p_{+1}^{(i)}-p_{-1}^{\left( i\right) }$, which reduces the two-state master equation for each of the elements in the network to the scalar rate equations
\begin{equation}
\frac{d\Pi ^{\left( i\right) }}{dt}=2g_{0}\sinh \left( K\Pi \right)
-2g_{0}\Pi ^{\left( i\right) }\cosh \left( K\Pi \right) .  \label{TSME}
\end{equation}%
Alternatively we may introduce the global difference variable $\Pi=p_{+1}-p_{-1}$ and replace the set of two-state master equations with the
single expression for the mean field variable%
\begin{equation}
\frac{d\Pi }{dt}=2g_{0}\sinh \left( K\Pi \right) -2g_{0}\Pi \cosh \left(
K\Pi \right) =-\frac{\partial V\left( \Pi \right) }{\partial \Pi }.
\label{global master}
\end{equation}%
The global two-state master equation (Eq. \ref{global master}) is equivalent to the over-damped movement of a particle in the potential
\begin{equation}
V\left( \Pi \right) \equiv \frac{2g_{0}}{K}\left[ \Pi \sinh \left( K\Pi
\right) -\left( 1+\frac{1}{K}\right) \cosh \left( K\Pi \right) \right] ,
\label{potential}
\end{equation}%
whose shape is a function of the cooperation parameter $K$ \cite{bianco08,turalska09}. For $K<1$, $V(\Pi )$ has one minimum centered at $\Pi=0$, denoting the fact that the equilibrium solution to Eq. \ref{global master} in this regime is $\Pi _{eq}=0$ and exactly half of the individuals in the network are in state $+1$ and half in the state $-1$. At the critical value $K_{C}=1$ a bifurcation occurs and the potential develops two wells separated by a barrier. The height of the barrier increases with an increase of the cooperation parameter $K$ as depicted in Fig. \ref{fig_well}. The presence of two equilibrium values $\Pi _{eq}(K)$, which correspond to the minima of the double well potential, reflects the fact that for $K>K_{C}$ a majority decision in the system emerges, where more than half of the individuals share the same state. The condition of perfect consensus is
reached for $K\rightarrow \infty $, where $\Pi $ converges on either the value $+1$ or $-1$, for details see \cite{turalska09,turalska11}.

\subsection*{Finite ATA\ Network}

In numerical calculations, the states of $N$ individuals composing the network, are first randomly initialized with the value of $+1$ or $-1$. Then, in a single time step the transition rates defined by Eq. \ref{rates} are calculated, according to which each individual is given the possibility to change its state. The procedure is repeated at all consecutive time steps. When $K=0$ all units are isolated and switch their state from $+1$ to $-1$ and in the opposite direction with the transition rate $g_{0}$. When the control parameter increases, $K>0$, a node in state $+1(-1)$ makes a transition to the state $-1(+1)$ slower or faster depending on whether the total number of nodes in state $+1$, $N_{+1}$ is larger or smaller than the total number of individuals in state $-1$, $N_{-1}$.

\begin{figure}[t]
\includegraphics{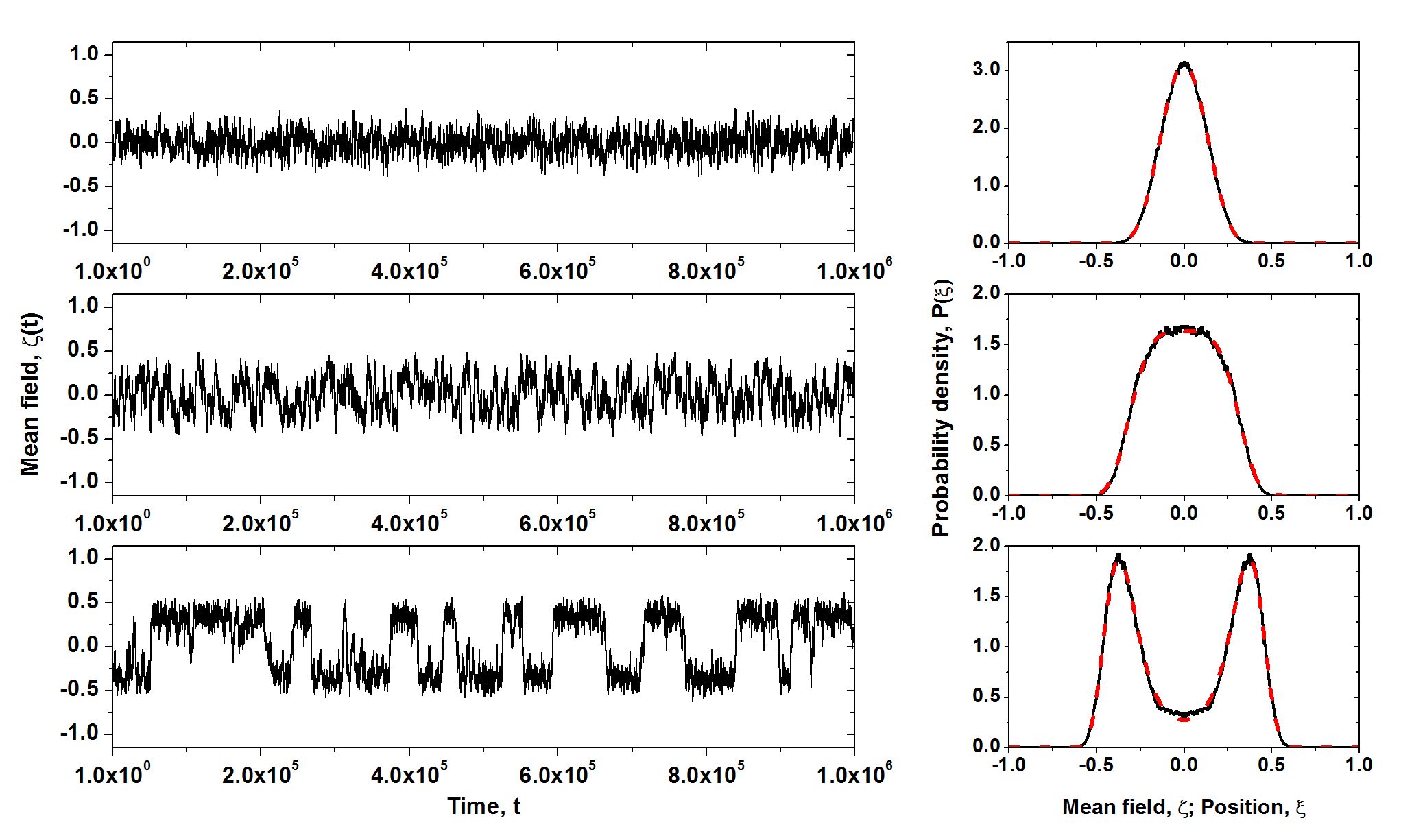}
\caption{{\bf Global order parameter $\protect\varsigma (K,t)$ evolution and corresponding probability distribution $P(\protect\xi )$.}
\textit{(left)} Temporal evolution of the mean field
variable $\protect\varsigma (K,t)$ for $K=0.95$, $K=1.00$ and $K=1.05$ is presented on upper, middle and lower panel, respectively. Network size is $N=1000$ nodes and $g_{0}=0.01$. \textit{(right)} Thin line depicts corresponding equilibrium probability distribution $P(\protect\varsigma )$. Dashed line denotes the best fit by Eq.(\protect\ref{equilibrium}).}
\label{fig_meanF}
\end{figure}

Global decisions of a network composed of $N$ individuals can be defined by the time-dependent global order parameter (mean field) $\varsigma (K,t)$ as
an average state of the network at a given time

\begin{equation}
\varsigma \left( K,t\right) =\frac{1}{N}\underset{j=1}{\overset{N}{\sum }}%
s_{j}\left( K,t\right)  \label{mean field}
\end{equation}%
where $s_{j}\left( K,t\right) $ is the value $(\pm 1)$ of the element $j$ of the network coupled to other nodes with cooperation level $K$. For values of the cooperation parameter $K<K_{C},$ single individuals are only weakly influenced by other individuals, leading to small amplitude fluctuations of $\varsigma (K,t)$, oscillating rapidly about the zero-axis [See Fig. \ref{fig_meanF}]. For $K>K_{C},$ the interaction between individuals gives rise to a majority state, which recovers partially the dichotomous character of the single individuals, however at much larger time scale.

The time-average of the global order parameter, $\varsigma _{eq}$ $\equiv\left\langle \left\vert \varsigma \left( K,t\right) \right\vert\right\rangle $, is used as a measure of the organization of the network and an approximation of its equilibrium state. In Fig. \ref{fig_well} the calculations of $\Pi _{eq}$ and $\varsigma _{eq}$ are compared and found to be essentially the same quantity, the difference being due to the finite number of elements in the numerical calculation. Thus, the global dynamics of the DMM is characterized by a phase transition with respect to the cooperation parameter $K$, demonstrating that a network of identical imitating individuals is able to reach consensus, given sufficient influence of the imitation on their decisions.

\subsection*{Langevin Formalism for Finite ATA\ Network}

In a network of finite size the mean field approximation is no longer exact and we replace $\Pi $ in Eq. \ref{global master} with
\begin{equation}
\frac{N_{+}\left( t\right) -N_{-}\left( t\right) }{N}=\xi \left( K,t\right)
+f\left( K,t\right) =\varsigma (K,t)
\end{equation}%
where $f\left( K,t\right) $ is a random fluctuation induced by the finite size of the network, whose magnitude is proportional to $1/\sqrt{N}$. When the network size increases, $N\rightarrow \infty $, the frequencies $N_{\pm1}(t)/N$ collapse into probabilities, $N_{\pm 1}(t)/N\rightarrow p_{\pm1}\left( t\right) $, and the global order parameter $\varsigma (K,t)=\xi(K,t)\rightarrow \Pi (t)$.

Using this new value of the global variable Eq. \ref{global master} becomes the nonlinear Langevin equation \cite{bianco08,turalska09}%
\begin{eqnarray}
\frac{d\xi }{dt} &=&2g_{0}\sinh \left( K\xi +Kf\right) -2g_{0}\left( \xi
+f\right) \cosh \left( K\xi +Kf\right)  \notag \\
&\approx &-\frac{\partial V\left( \xi \right) }{\partial \xi }+\eta (t)
\label{global Langevin}
\end{eqnarray}%
where $\eta (t)$ is a random function given by $\eta (t)=\sigma f(K,t)$ and the potential is given by Eq. \ref{potential}.

The strength $\sigma $ of the random force is determined by the equilibrium properties of the numerical realization of the DMM dynamics. We approximate the random force by a white noise Gaussian process whose strength is $\sigma=\sqrt{2D}$ and $D$ is an unknown diffusion coefficient. The Langevin equation given by Eq. \ref{global Langevin} has an equivalent Fokker-Planck equation (FPE) description%
\begin{equation}
\frac{\partial P\left( \xi ,t\right) }{\partial t}=\frac{\partial }{\partial
\xi }\left[ \frac{\partial V\left( \xi \right) }{\partial \xi }P\left( \xi
,t\right) +D\frac{\partial P\left( \xi ,t\right) }{\partial \xi }\right]
\label{FPE}
\end{equation}%
for the probability density function (PDF). The equilibrium solution to the FPE is obtained by setting the time derivative to zero to obtain the equilibrium distribution for the mean field variable

\begin{equation}
P_{eq}\left( \xi \right) =\frac{1}{Z}\exp \left[ -\frac{V\left( \xi \right)
}{D}\right]  \label{equilibrium}
\end{equation}%
and $Z$ is the normalization.

\begin{figure}[t]
\includegraphics[scale=1.50]{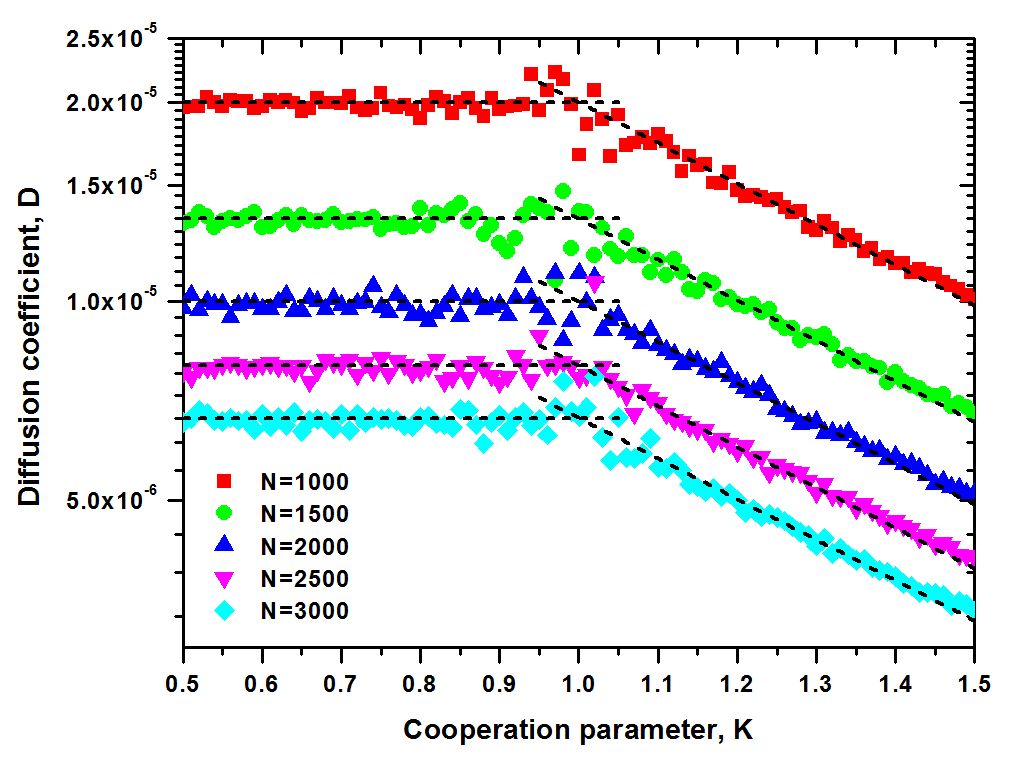}
\caption{{\bf Dependence of the diffusion coefficient $D$ on the cooperation parameter $K$ for all-to-all network of increasing size.}
Dashed lines represent analytic expression for $D$, Eq.(\protect\ref{analytic D}). Note that the vertical scale for the diffusion coefficient is logarithmic.}
\label{fig_diffusion}
\end{figure}

The value of $D$ is determined numerically from the mean field generated by the DMM of size $N$, with cooperation parameter $K$ and transition rate $g_{0}.$ Using a long time realization of $\varsigma \left( K,t\right),$ the equilibrium PDF can be estimated as depicted in Fig. \ref{fig_meanF}. The resulting histograms are fit with Eq. \ref{equilibrium}, where $D$ is kept as an unknown parameter that optimizes the fitting. Fig. \ref{fig_diffusion} presents the dependence of numerically obtained values of $D$ on the cooperation parameter $K$ and size of the network $N.$ It is well known from the law of large numbers that the intensity of the noise ought to decrease as $N^{-1/2}$. Using this as a starting point we obtain an empirical expression defining the dependence of $D$ on $K$ and $N$:%
\begin{equation}
D=\left\{
\begin{array}{l}
\frac{2g_{0}}{N} \\
\frac{2g_{0}}{N}\exp \left[ -\sqrt{2}\left( K-1\right) \right]%
\end{array}%
\right.
\begin{array}{cc}
& K<K_{C} \\
& K>K_{C}%
\end{array}
\label{analytic D}
\end{equation}%
whose excellent overlap with values of $D$ in two domains, $K<K_{C}$, where the value of $D$ is constant and independent of the cooperation parameter, and $K>K_{C}$, where $D$ decreases exponentially with $K$, as depicted in Fig. \ref{fig_diffusion}.

\begin{figure}[t]
\includegraphics{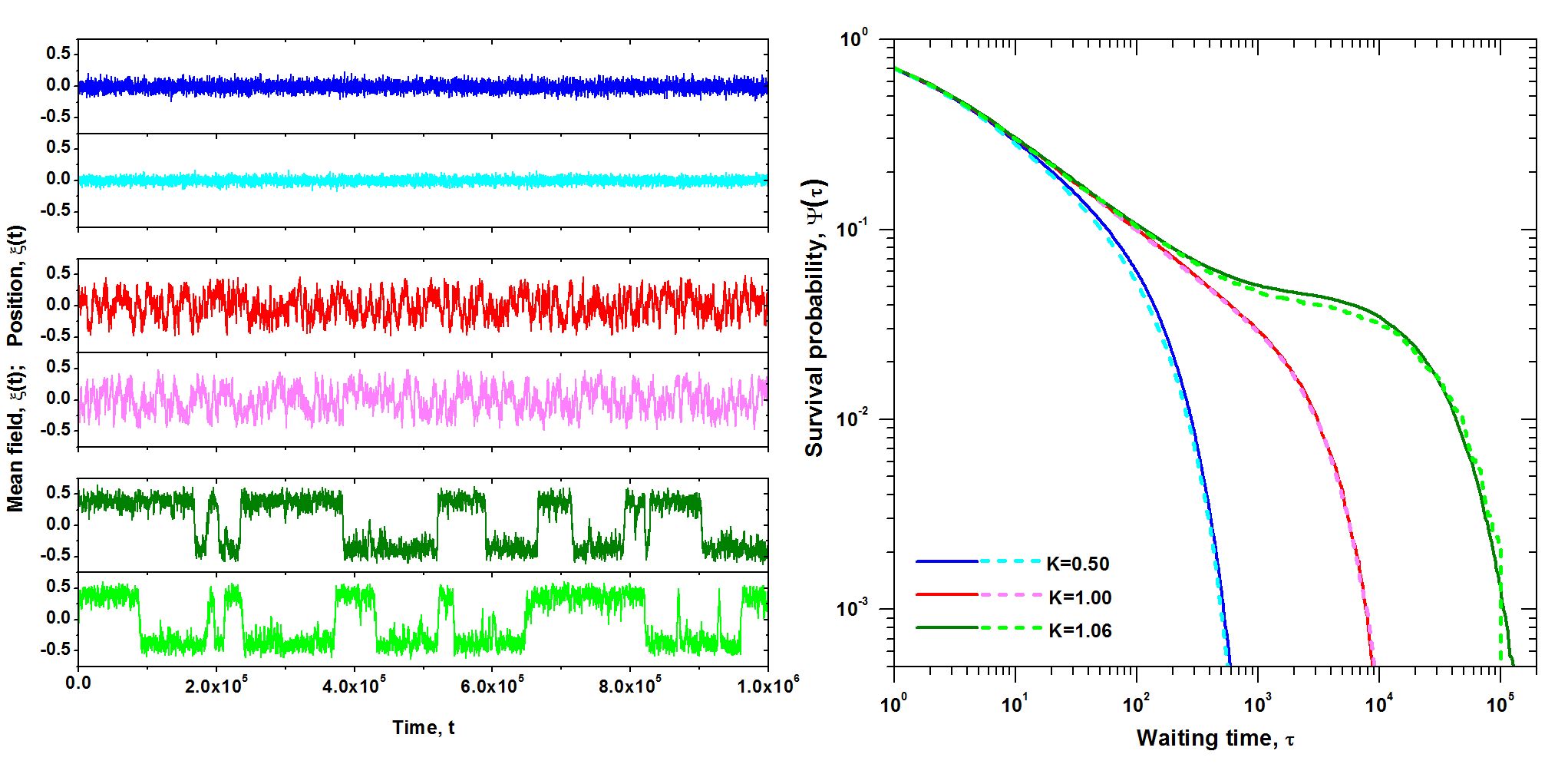}
\caption{{\bf Comparison of the global behavior of an all-to-all network of $N=1000$ nodes with the solution to the corresponding Langevin equation.}
\textit{(left)} Temporal evolution of mean field variable (top part of each panel) is compared with the numerical integration of Eq.(\protect\ref{global Langevin}) (bottom part of each panel).\textit{\ }Adopted values of cooperation parameter are $K=0.50$, $K=1.00$ and $K=1.06$ for the top, middle and bottom panels, respectively. \textit{\ (right)} The comparison of the survival probabilities of time intervals between consecutive zero crossings for both variables. Straight and dashed lines correspond to the DMM and the Langevin case, respectively.}
\label{fig_compare}
\end{figure}

Now one can adopt the Langevin equation (Eq. \ref{global Langevin}) to quantitatively simulate the dynamics of the DMM\ on a finite size ATA network. For given $N$, $K$ and $g_{0}$ Eq. \ref{global Langevin} is numerically integrated using the Euler-Maruyama algorithm for integration of stochastic differential equations. The simulation time step is $h=0.01$. The comparison of the fluctuations of the global order parameter $\varsigma(K,t),$ resulting from the DMM\ dynamics defined by Eq. \ref{rates}, and the fluctuations of the mean field $\xi (K,t)$, resulting from the solution to the Langevin equation with matching parameters is presented in Fig. \ref{fig_compare}. The visual similarity of both time series is further confirmed by the comparison of their statistical properties. We quantify the changes in temporal properties of $\varsigma (K,t)$ and $\xi (K,t)$\ by evaluating the waiting time PDF $\psi (\tau )$ and corresponding survival probability function,
\begin{equation}
\Psi (\tau )=\int_{\tau }^{\infty }\psi (t^{\prime })dt^{\prime },
\label{survival PF}
\end{equation}%
of time intervals $\tau $ between consecutive crossings of the zero-axis by $\varsigma (K,t)$ and $\xi (K,t)$. The statistical properties of the global order parameter $\varsigma (K,t)$ have been extensively studied for subcritical $(K<K_{C})$, critical $(K\approx K_{C})$ and supercritical $(K>K_{C})$ regime of values of the cooperation parameter \cite{turalska09,west14}. Briefly, in the subcritical regime $\Psi (\tau )$ has an exponential form, which reflects the largely independent nature of individuals. The critical region is characterized by an inverse power law decrease of $\Psi (\tau )$,\ whereas in the supercritical region long majority intervals are responsible for an exponential shoulder present in $\Psi (\tau ).$ From the comparison of respective survival probability functions [See Fig. \ref{fig_compare}] it is evident that the Langevin equation reproduces all the global dynamic properties of the DMM network dynamics, throughout the range of values of the cooperation parameter $K.$

\section*{Results \label{network response}}

In this section we consider the interaction between two ATA DMM networks each of which is modeled by a Langevin equation with a double-well potential. The symmetry of the response of the combined network depends on the choice of coupling between the sub-networks, which we explain in detail. Section \ref{adam} presents a approximation technique for modeling the coupling between the two networks through a modification of the individual switching rates in terms of the fraction of individuals within a network that are sensitive to the state of the other network. It is shown in Section \ref{two} that the sign of the perturbative coupling between the two networks leads to either synchronization or anti-synchronization of their relative dynamics. The relative entropy is used to determine that the conditions under which the maximum information is exchanged between the two
networks. The limitation of the coupled Langevin model as an approximation to the coupling between two ATA DMM networks is also discussed.

\subsection*{Network Response \label{adam}}

The response of the ATA DMM network to external forces was analyzed by Svenkeson \cite{svenkeson13} and we follow that discussion closely in this section. We assume that a subset of individuals within the group are sensitive to external influences and that they relay the information obtained through this sensitivity to the \textit{free individuals} in the network (free from external influences). These latter individuals behave in the normal DMM fashion previously described. The total number of individuals in the network $N$ is the sum of those that are free $n$ and those that are driven by the external force $l$, $N=n+l.$ Consequently, in the presence of sensitive individuals the transition rates governing the ATA network dynamics of the free individuals are

\begin{eqnarray}
g_{_{+-}}(t) &=&g_{0}\exp \left[ -K\left[ \left( 1-P\right) x+Ps_{l}\right] %
\right] ;  \notag \\
g_{_{-+}}(t) &=&g_{0}\exp \left[ K\left[ \left( 1-P\right) x+Ps_{l}\right] %
\right]   \label{perturbed}
\end{eqnarray}%
where the fraction of sensitive individuals is $P=\frac{l}{N}$ and the fraction of free individuals is $1-P=\frac{n}{N}.$ The mean field for the free individuals alone is given by%
\begin{equation}
x\left( K,t\right) \equiv \frac{1}{n}\overset{n}{\underset{j=1}{\sum }}%
s_{j}\left( K,t\right) =\frac{n_{+}\left( K,t\right) -n_{-}\left( K,t\right)
}{n}.  \label{new mean field}
\end{equation}%
The two-state master equation for the mean field coupled to an external force is therefore given by%
\begin{eqnarray}
\frac{d\Pi }{dt} &=&2g_{0}\left[ \sinh \left( K_{P}x\right) -\Pi \cosh
\left( K_{P}x\right) \right] \cosh \left( KPs_{l}\right)   \notag \\
&&+2g_{0}\left[ \cosh \left( K_{P}x\right) -\Pi \sinh \left( K_{P}x\right) %
\right] \sinh \left( KPs_{l}\right)
\end{eqnarray}%
with the new control parameter $K_{P}\equiv \left( 1-P\right) K.$

\subsection*{Two Coupled Networks \label{two}}

\begin{figure}[t]
\includegraphics{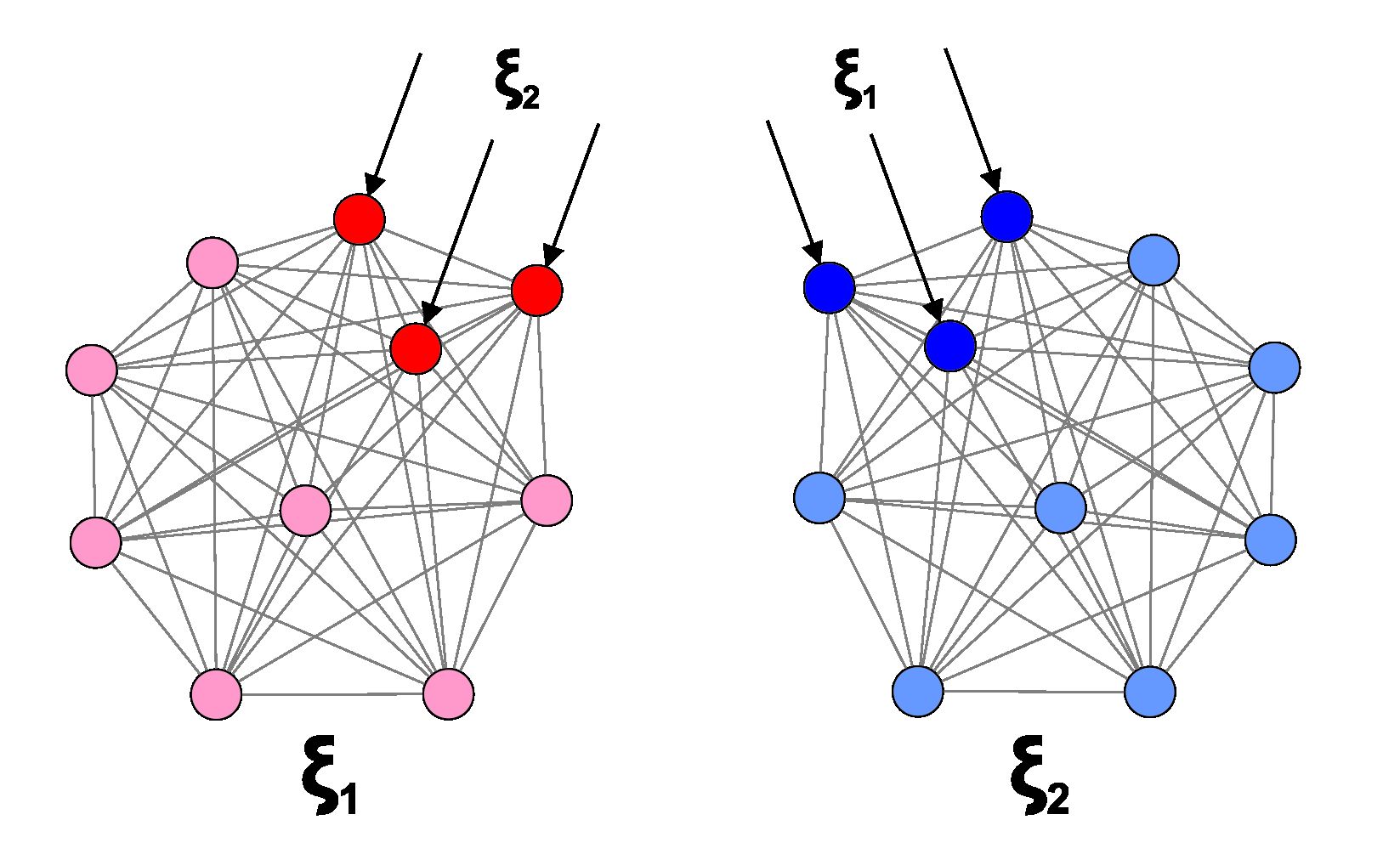}
\caption{{\bf Coupling between two ATA networks is realized by letting a fraction of nodes of each network sense the mean field of the other network.}}
\label{fig_network}
\end{figure}

Here we extend the discussion of the network response to external forces to the case when the external influence originates from another ATA\ network. Additionally, we assume that the second ATA network is in turn being influenced by the very network it exerts its influence on, thus leading to bi-directional coupling of two ATA networks [See Fig. \ref{fig_network}]. Taking advantage of the demonstrated equivalence between the dynamics of the DMM\ on an ATA\ network and the solution to the matching Langevin equation, we study the coupling of two ATA networks by means of the coupled Langevin equations

\begin{eqnarray}
\frac{d\xi _{1}}{dt} &=&-\frac{\partial V\left( \xi _{1}\right) }{\partial
\xi _{1}}+\eta _{1}(t)\pm \alpha \xi _{2}  \notag \\
\frac{d\xi _{2}}{dt} &=&-\frac{\partial V\left( \xi _{2}\right) }{\partial
\xi _{2}}+\eta _{2}(t)\pm \alpha \xi _{1}  \label{coupled}
\end{eqnarray}%
where $\alpha $ is the strength of the coupling between networks and the choice of sign determines the symmetry of the interaction. Thus, in the model one considers the intra-network coupling between elements of single ATA network, the cooperation parameter $K,$ and the inter-network coupling $\pm \alpha $, denoting the mutual influence between networks.

\subsubsection*{Symmetry of interaction}

\begin{figure}[t]
\includegraphics[scale=2.00]{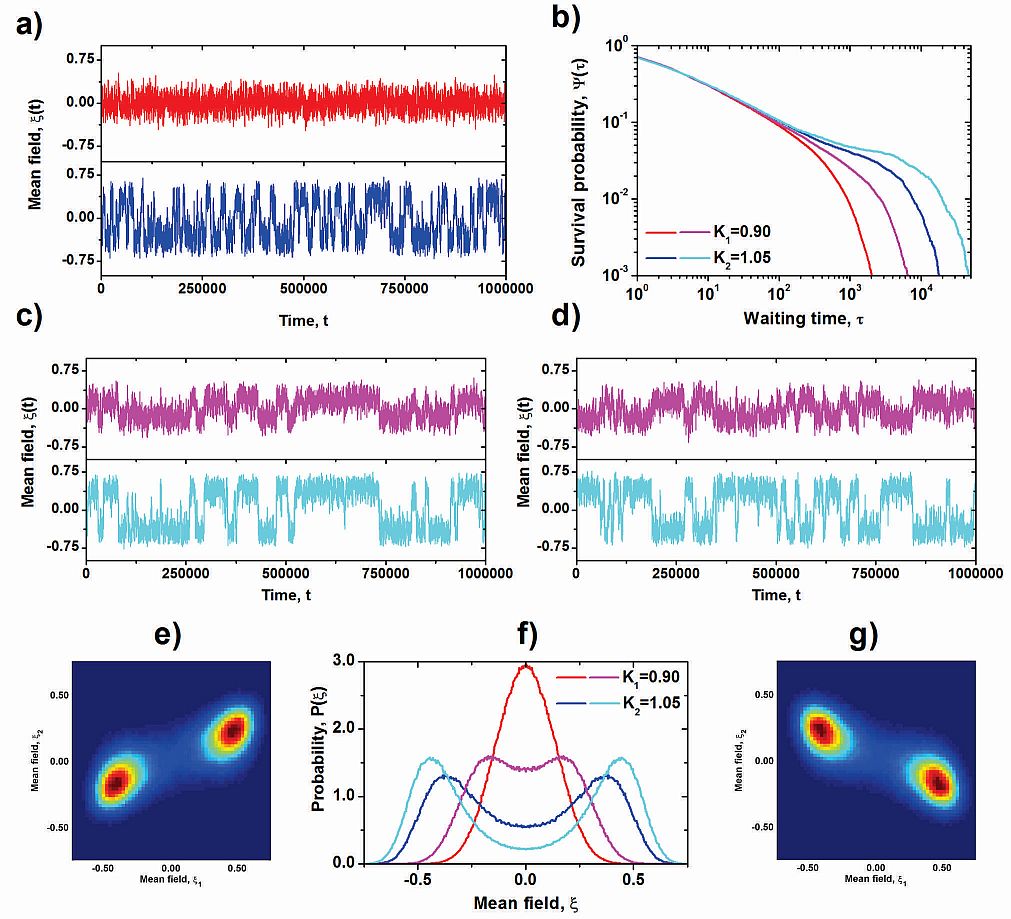}
\caption{{\bf The coupling of two Langevin equations.}
The cooperation parameters are $K_{1}=0.90$ and $K_{2}=1.05$ for the two networks, and $\protect\alpha =0.001,$ with positive sign \textit{(left)}, with negative sign \textit{(right)}. A) The global variable fluctuations for the uncoupled case. Top panel (blue line) corresponds to $K_{1}$, while bottom panel (red line) corresponds to $K_{2}.$ B) The distribution of waiting times demonstrates the change introduced by coupling to the dynamics of both networks. C) Synchronous coupling of two networks. D)\ Antisynchronous coupling of two networks. In both cases top panel (light blue) corresponds to system $K_{1}$ and bottom panel (pink) shows the behavior of system $K_{2}.$ E, G) The joint probability distribution for the coupled variables $P(\protect\xi _{1},\protect\xi _{2}).$ F) Comparison of single variable probability distribution between the coupled and uncoupled cases. Synchronous and antisynchronous coupling of the same strength result in the same probability distributions.}
\label{fig_synch_anti}
\end{figure}

The positive sign for the interaction terms in Eq. \ref{coupled} leads to a synchronized response of the two networks. This is observed in Fig. \ref{fig_synch_anti} where the temporal fluctuations of the mean field $\xi_{1}(K_{1},t)$ and $\xi _{2}(K_{2},t)$ and their statistical properties are compared between uncoupled $\left( \alpha =0\right) $ and coupled case. In the uncoupled case (Fig. \ref{fig_synch_anti}A) the disparity between the the behavior of $\xi _{1}(K_{1},t)$ and $\xi _{2}(K_{2},t)$ is clearly visible, since the subcritical system is characterized by random fluctuations, while the supercritical system demonstrates the onset of dichotomous dynamics. When positive coupling is turned on, it is evident that although not in lock step the two networks are synchronized (Fig. \ref{fig_synch_anti}C). Variable $\xi _{1}(K_{1},t)$ is being visibly influenced
by coupling to $\xi _{2}(K_{2},t)$, exhibiting dichotomous dynamics, and although it does not match $\xi _{2}(K_{2},t)$ in amplitude, it follows the switching dynamics of $\xi _{2}(K_{2},t).$ The joint probability distribution for both variables, $P\left( \xi _{1},\xi _{2}\right) $, provides additional evidence for the synchronization between systems, being characterized by two peaks, corresponding to the situation when both global variables simultaneously attain positive or negative values. The coupling effects are visible also in the effective increase of cooperation parameters $K$ for both networks. When compared with the uncoupled case, the probability distributions of the global variables, $P(\xi _{1})$ and $P(\xi _{2})$, demonstrate stronger bimodality, especially evident for the $\xi_{1}(K_{1},t)$ variable, whose original distribution is unimodal (Fig. \ref{fig_synch_anti}F).

The negative sign for the interaction terms in Eq. \ref{coupled} leads to an anti-synchronized mutual response of the two networks, observed in Fig. \ref{fig_synch_anti}D. It is again evident that although not in lock step the two networks are negatively synchronized. The probability distribution of global variable $P(\xi _{1})$ is clearly unimodal in the absence of coupling and bimodal when the mutual interaction is turned on (Fig. \ref{fig_synch_anti}F). The $P(\xi _{2})$ is affected less significantly, with the bimodal peaks being shifted towards higher values.of $\xi _{2}$ as a result of the coupling. Due to the the fact that the coupling $\alpha$ differs only in sign between the synchronized and anti-synchronized case, the obtained probability distributions are indistinguishable. The presence of the bimodality is clearly indicated in the shape of distribution $P\left(
\xi _{1},\xi _{2}\right) $ (Fig. \ref{fig_synch_anti}G) and here is where the effect of the sign of the mutual coupling is most evident. The anti-synchronization is manifest by the two networks peaking oppositely, when one global variable is positive the other is negative and vice versa.

Let us look in more detail at the reason for the anti-synchronization.

\subsubsection*{Anti-synchronization}

In numerous studies \cite{pikovsky} the coupling of stochastic oscillators represented by the Langevin equations leads to network size resonance, when the ensemble of oscillators respond to a periodic forcing in a fashion similar to the stochastic resonance phenomenon. In our case however, the weak coupling of two Langevin equations results in non-resonant behavior. In particular, the coupling causes an effective increase of the cooperation parameter $K$ in both networks, regardless of the sign of the coupling.

A visual inspection of the temporal fluctuations of the mean fields $\xi_{1}(K_{1},t)$ and $\xi _{2}(K_{2},t)$ presented on Fig. \ref{fig_synch_anti}D demonstrates the change in the dichotomous character of both networks, with the increased sojourn in the majority states visible in $\xi _{1}(K_{1},t)$ and $\xi _{2}(K_{2},t),$ when compared to the uncoupled dynamics (Fig. \ref{fig_synch_anti}A). Longer waiting times induced by the mutual coupling affect the survival probability function as well. In Fig. \ref{fig_synch_anti}B we see that the inverse power-law regime of the survival probability is not changed. The difference in $\Psi (\tau )$ due to the coupling increases the extension of the exponential shoulder, which in an uncoupled case would correspond to a larger cooperation parameter $K$. Thus, individuals present in each of the two coupled ATA\ networks perceive
the coupling as an effective increase of the level of intra-network cooperation. Since the sign of the coupling $\alpha $ changes only the nature of synchronization, the synchronized case presents the same change to the survival probability functions as the anti-synchronized coupling.

\begin{figure}[h]
\includegraphics[scale=1.20]{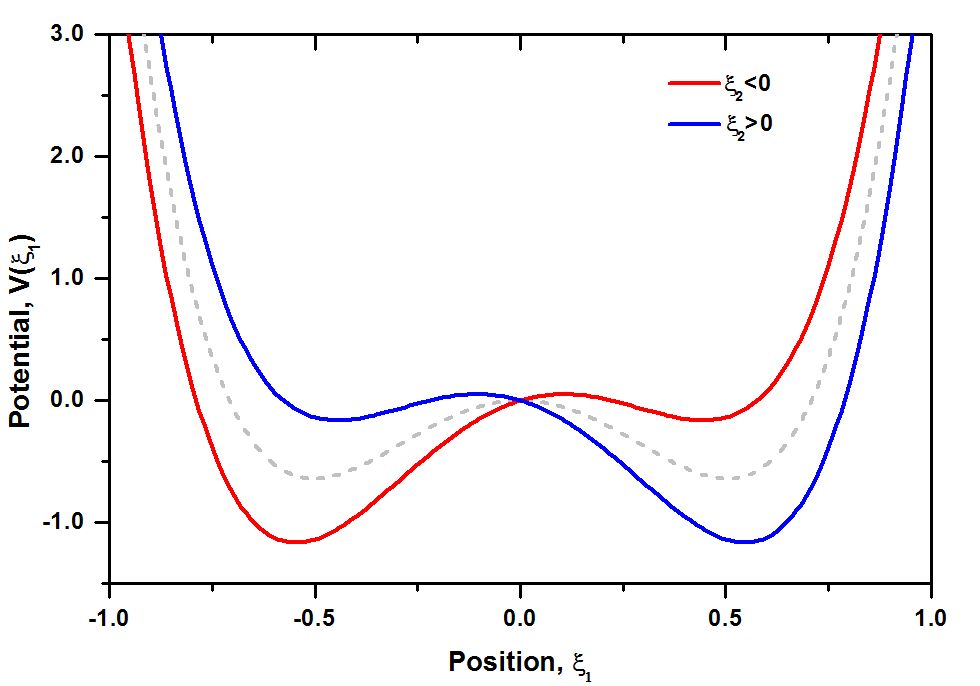}
\caption{{\bf Symmetric double well potential of the uncoupled networks (dashed curve) become mutually biased when weakly coupled to one another (see discussion in text).}}
\label{fig_potential}
\end{figure}

Of most significance is the observation that the two time series are statistically anti-synchronized. Consequently, when the mean field $\xi_{1}(K_{1},t)$ is in the positive state the probability is greatly enhanced that the mean field $\xi _{2}(K_{2},t)$ is in the negative state and vice versa. The anti-synchrony effect can be understood in the following way. Consider the potential for network one with the interaction term $V\left(\xi _{1}\right) +\alpha \xi _{1}\xi _{2}.$ When $\alpha =0$ the potential is symmetric around $\xi _{1}=0$ and we have the uncoupled case of the network being in either consensus state with equal frequency. However when $\alpha\neq 0$ and $\xi _{2}<0$ the network one potential becomes asymmetric with the deeper well being on the side $\xi _{1}>0.$ This biasing of the potential [See Fig. \ref{fig_potential}] results in the mean field time series for network one being in the positive state. In this case of $\xi_{1}>0$ consider the potential for network two $V\left( \xi _{2}\right)+\alpha \xi _{1}\xi _{2}$. This latter potential is biased by the mean field of network one and is now also asymmetric. With $\xi _{1}>0$ the deeper well in network two is for $\xi _{2}<0\ $resulting in the mean field for network two being in the negative consensus state. However the two networks are not locked in these relative states. The fluctuations are seen to be substantial and when they are sufficiently large to induce a phase transition in one network the inter-network coupling induces a corresponding change in the dynamics of the other network. This intermittent rocking of the two potentials produces the anti-symmetry of the two time series observed in Fig. \ref{fig_synch_anti}.

\subsection*{Information Exchange}

A topic of considerable interest are the parameter values at which the maximum information exchange between the interacting systems takes place. One measure of this exchange is the relative entropy $S$ defined here in terms of the mean fields of the two networks%
\begin{equation}
S=S_{0}-\int P\left( \xi _{1}^{CP},\xi _{2}^{CP}\right) \log _{2}\left[
\frac{P\left( \xi _{1}^{CP},\xi _{2}^{CP}\right) }{P\left( \xi
_{1}^{UC}\right) P\left( \xi _{2}^{UC}\right) }\right] d\xi _{1}d\xi _{2}.
\label{entropy}
\end{equation}%
The relative entropy relates the entropy of the coupled networks, expressed by the joint probability distribution of mean fields in the coupled case, $P\left( \xi _{1}^{CP},\xi _{2}^{CP}\right) $, to the entropies of the individual uncoupled networks, denoted by probability distributions for uncoupled variables, $P\left( \xi _{1}^{UC}\right) $ and $P\left( \xi_{2}^{UC}\right) $. The constant $S_{0}$ denotes a reference state.

\begin{figure}[h]
\includegraphics[scale=1.20]{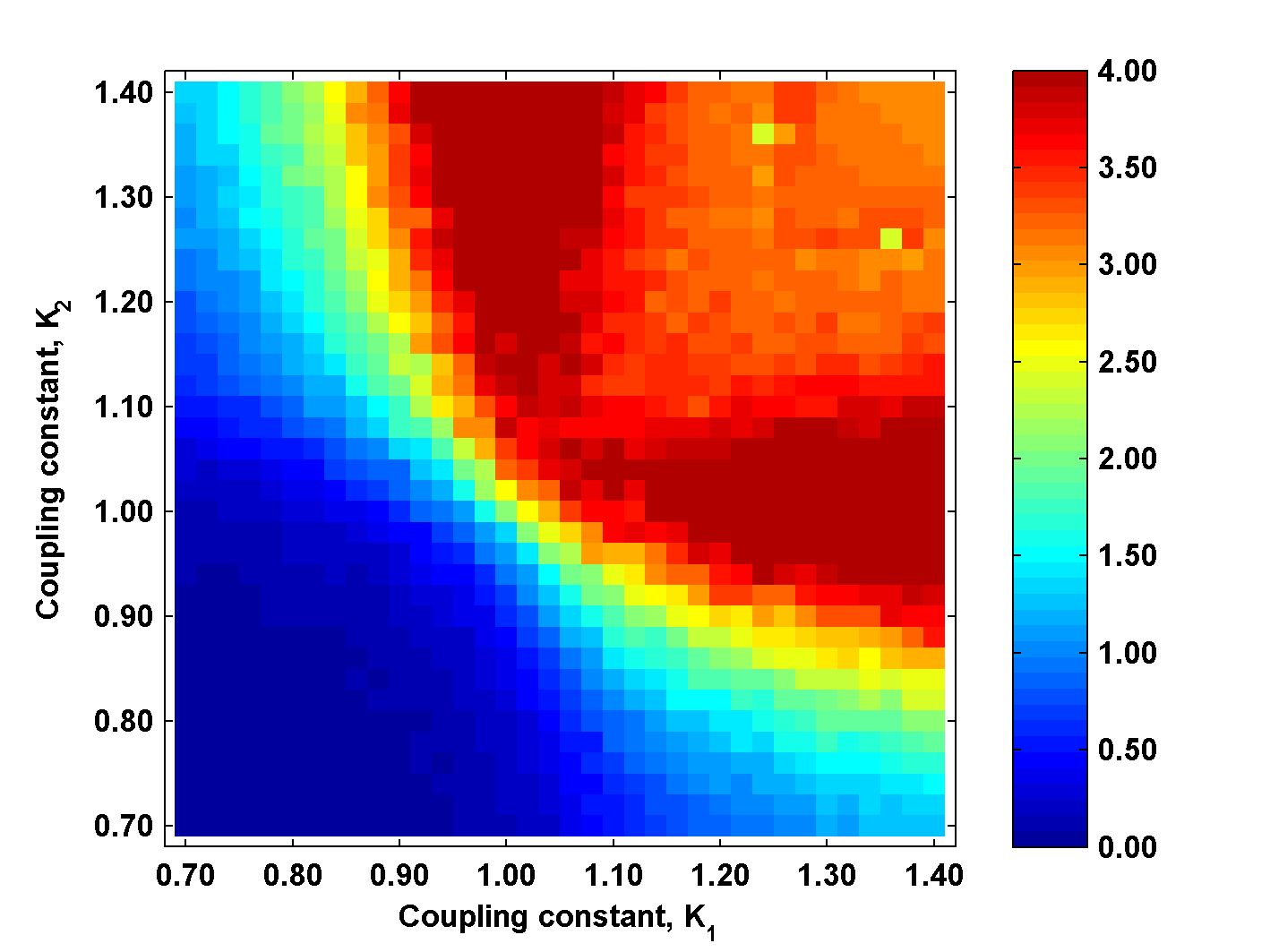}
\caption{{\bf The values of relative entropy $S(K_{1},K_{2})$ of Eq.(\protect\ref{entropy}) are color coded for each pair of cooperation parameters $K_{1}$ and $K_{2}$.}
Coupled Langevin equations were numerically integrated with $N=500$ and $g_{0}=0.01$. The coupling strength is $\protect\alpha =0.001$.}
\label{fig_map}
\end{figure}

The values of the relative entropy calculated for the coupled Langevin equations model with a wide range of cooperation parameters $K_{1}$ (for $\xi _{1}(K_{1},t)$ variable) and $K_{2}$ (for $\xi _{2}(K_{2},t)$ variable) are shown on Fig. \ref{fig_map}. The mirror symmetry of $S(K_{1},K_{2})$ with respect to the $K_{1}=K_{2}$ line is due to the symmetric nature of the coupling between the two Langevin equations in terms of the PDF's. Additionally, as demonstrated on Fig. \ref{fig_synch_anti}F, since the PDF's do not depend on the sign of the coupling between two networks, the values of relative entropy are identical both for synchronous and anti-synchronous coupling.

The concentration of highest values of $S(K_{1},K_{2})$ near $K_{1}\approx 1$ clearly indicates that maximum transfer of information between the two networks occurs at the critical value of the cooperation parameter $K_{1};$ an observation consistent with the PCM \cite{west08}. In the regime where both cooperation parameters are subcritical, $K_{1},K_{2}<K_{C}$, the relative entropy $S(K_{1},K_{2})$ values are small, implying a lack of effective coupling between the two networks. When $K_{1},K_{2}>K_{C}$, the values of $S(K_{1},K_{2})$ are relatively high, although smaller then maximal, indicating that the intra-network dynamical organization facilitates the cooperation between networks. It is worth pointing out that the peak in value of the relative entropy disappears when both $K_{1}=K_{2}=1,$ indicating that there is no preference for information transfer when both networks are at criticality.

\begin{figure}[h]
\includegraphics{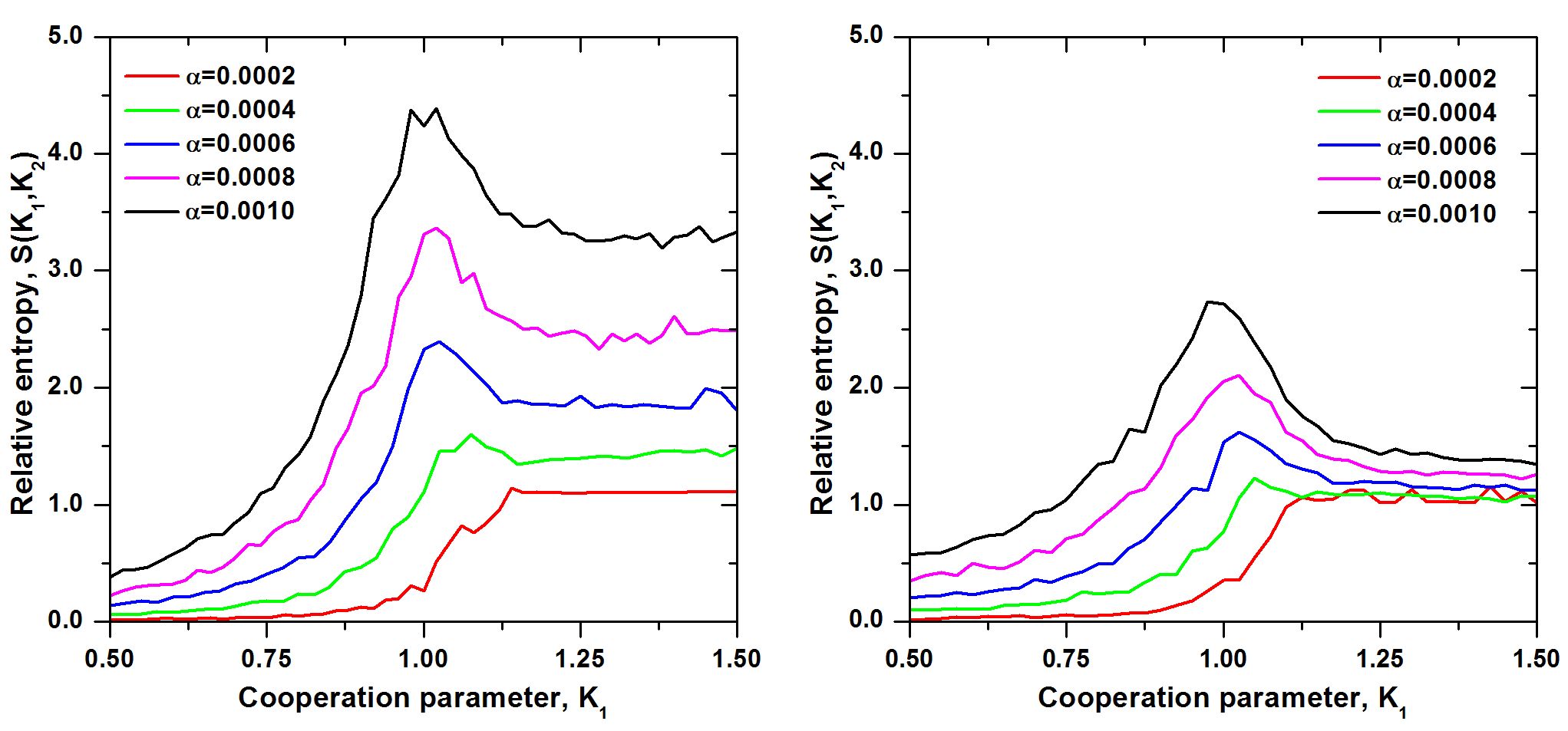}
\caption{{\bf Coupling of two Langevin equations is compared with coupling of two all-to-all networks.}
 Values of the relative entropy for increasing strength of coupling $\protect\alpha $
between two Langevin equations \textit{(left)} are compared with values of the relative entropy obtained for two connected networks \textit{(right)}. In both cases $N=500$ and $g_{0}=0.01$ and $K_{2}=1.20.$ The sizes of the fraction of sensitive nodes adopted for the coupling of two ATA\ networks were increasing from $P=0.01$ to $P=0.05$ for increasing values of $\protect\alpha.$}
\label{fig_entropy_cut}
\end{figure}

Furthermore, the effective influence of the coupling between networks depends on the strength of the coupling term. The values of relative entropy obtained for increasing the coupling strength $\alpha $ reinforce the observation that the transfer of information is maximal once one network is in the critical regime [See Fig. \ref{fig_entropy_cut}]. The values of $S$ initially increase with the increase of cooperation parameter $K_{1},$ to reach maximum at $K_{1}\approx 1$. This peak is followed by a decrease of $S$ and a region of $K_{1}$ in which the relative entropy is constant suggesting that the two-point PDF is proportional to the product of the singlet PDF's, $P\left( \xi _{1}^{CP},\xi _{2}^{CP}\right) =P\left( \xi _{1}^{CP}\right) P\left( \xi _{2}^{CP}\right) $. If this ratio is denoted by $C$, then the change in the relative entropy is given by%
\begin{equation}
\Delta S=-C\log _{2}C  \label{entropy change}
\end{equation}%
and the constant $C$ is a function of the control parameters $C(K_{1},K_{2}). $ In addition to the dependence on the control parameters through $C$ the relative entropy is seen to decrease disproportionately with the coupling strength.

\subsection*{Validity of Langevin Formalism}

The discussion of the external influence exerted on the network through a small fraction of its members presented in Section \ref{adam} can be adopted to treat the weak coupling between two networks realized with the Langevin equation model. In that case one considers two ATA networks of size $N_{1}$ and $N_{2}$, whose intra-network dynamics is realized with the cooperation levels $K_{1}$ and $K_{2}$, respectively. Here for simplicity we choose $N_{1}=N_{2}$. Next the dynamics of a fraction $1-P$ of the free individuals is realized following the rules of normal DMM\ dynamics, adopting the expressions for the transition rates defined by Eq. \ref{rates}. The fraction $P$ of sensitive individuals obey the modified transition rates

\begin{eqnarray}
g_{_{+1\rightarrow -1}}^{\left( i\right) }(t) &=&g_{0}\exp \left[
-K_{1}\varsigma _{1}+\varsigma _{2}\right] ;  \label{modified rates} \\
g_{_{-1\rightarrow +1}}^{\left( i\right) }(t) &=&g_{0}\exp \left[
K_{1}\varsigma _{1}-\varsigma _{2}\right]
\end{eqnarray}%
where the additional term in the exponent comes from the coupling to the second network. The transition rates defined for the dynamics of the sensitive individuals in the second network contain opposite indices. With the introduction of the difference variable as before, $\Pi ^{\left(i\right) }=p_{+1}^{(i)}-p_{-1}^{\left( i\right) }$, the $1-P$\ two-state master equations describing the free individuals reduce to Eq. \ref{TSME} and the $P$ sensitive individuals are described by

\begin{equation}
\frac{d\Pi ^{\left( i\right) }}{dt}=2g_{0}\sinh \left( K_{1}\Pi \right)
-2g_{0}\Pi ^{\left( i\right) }\cosh \left( K_{1}\Pi \right) -2g_{0}\varsigma
_{2}.  \label{TSME sens}
\end{equation}

The dynamics of the global variable for one network is then

\begin{equation*}
\frac{d\varsigma _{1}}{dt}=\frac{1}{N}\sum_{i=1}^{N}\frac{d\Pi ^{\left(
i\right) }}{dt}\approx -\frac{\partial V(\varsigma _{1})}{\partial \varsigma
_{1}}+\eta _{1}(t)-2g_{0}P\varsigma _{2}(K_{2},t)
\end{equation*}%
The correspondence with the coupled Langevin equation occurs when the coupling is $\alpha =2g_{0}P$, denoting the fact that in a particular realization weak coupling between two networks can be obtained by making a small fraction of individuals of each network sensitive to the dynamics of the second network.

The effect of the coupling of two ATA\ networks realized through sensitive individuals is demonstrated in Fig. \ref{fig_entropy_cut}, where the values of the relative entropy for increasing values of the coupling strength $\alpha $ are compared with the Langevin equation model. We observe the same qualitative behavior as in the Langevin models, that is, the maximum information transfer occurs at criticality, and the values of the entropy saturate in the supercritical cooperation parameter regime.

\begin{figure}[h]
\includegraphics{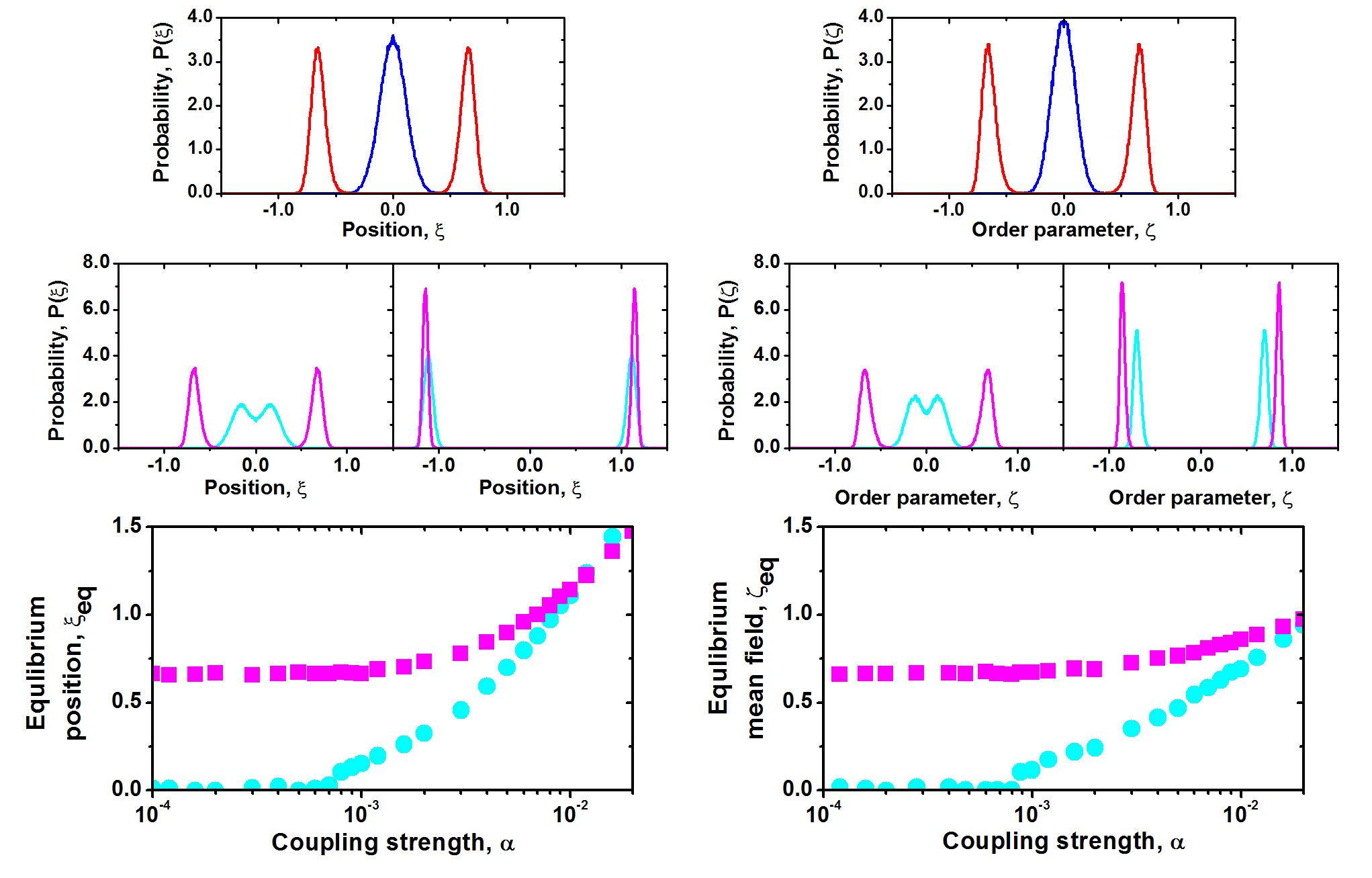}
\caption{{\bf The coupling of two networks results in an effective increase of the cooperation parameter characterizing the dynamics of each network.}
This effect can be observed by the shift in location of the maxima in the PDF for the mean field variable. \textit{(left)} Coupling of the two Langevin equations and \textit{(right)} coupling of two ATA networks is considered. In both cases top panels demonstrate the PDF's for uncoupled networks. Middle panels demonstrate the effect of coupling, $\protect\alpha =0.001$ (left middle panel) and $\protect\alpha=0.01$ (right middle panel), on the shape of the PDF. Bottom panels show the dependence of the equilibrium value of the mean field as a function of the coupling strength $\protect\alpha $. The size of both networks is $N=500$ and $g_{0}=0.01.$ The cooperation parameters are $K_{1}=0.80$ (blue and light blue lines) and $K_{2}=1.20$ (red and pink lines).}
\label{fig_shift}
\end{figure}

Finally, one needs to consider what is the range of validity in which the coupled Langevin model recovers the dynamics of two ATA\ networks coupled with a small number of links. One significant difference between the direct numerical simulation of an ATA\ network with DMM\ dynamics and the realization of the corresponding Langevin equation is the fact that in the DMM the mean field variable $\varsigma (K,t)$ is always limited to the range $\pm 1$, while the variable $\xi (K,t)$ present in the Langevin equation is not so restricted. Thus, we adopted the value of $\alpha $ at which $\xi(K,t)$ observed for the coupled networks is larger than either $+1$ or $-1$ as the measure of the validity limitation for the equivalence of the two approaches.

We investigate the properties of the PDF of the Langevin variables $\xi_{1}(K_{1},t)$ and $\xi _{2}(K_{2},t)$, and that of the DMM\ variables, $\varsigma _{1}(K_{1},t)$ and $\varsigma _{2}(K_{2},t)$. We concentrate our attention on the equilibrium value of $\xi _{1,2}$ and $\varsigma _{1,2}$, which corresponds to the location of the maxima of the PDF's for those variables. The approach is illustrated on Fig. \ref{fig_shift}, where the PDF's obtained for weak coupling, $\alpha =0.001$ are compared with those obtained for stronger coupling, $\alpha =0.01$. In the case of weak coupling both approaches lead to identical PDF's. However, stronger coupling in the case of coupled Langevin equations results in the peaks of $P(\xi _{1})$ and $P(\xi _{1})$ to be located outside the range $-1<\xi <1$, which violates the underlying assumption for the mean field dynamics. This behavior is not
observed for coupled ATA\ networks independently of the strength of the coupling.

The position of the equilibrium values of $\xi _{1,2}$ and $\varsigma _{1,2}$ for a wide range of $\alpha $ values is also plotted. One determines for coupling strengths $\alpha <10^{-3}$ that the coupling does not result in any effective change in either network, being too weak to influence the intra-network dynamics. Additionally, coupling values larger than $\alpha>10^{-2}$ do not recover the correspondence between the direct simulation of the DMM\ and the Langevin approach.

\section*{Discussion and Conclusions \label{discussion}}

Herein we have shown that two weakly coupled ATA DMM networks each modeled with a two-state master equation can be approximated to a high degree of accuracy by two Langevin equations each with a double well potential that is biased by the interaction with the other network. When the percentage of elements within the network that is sensitive to the other network is sufficiently small the coupling can be treated as a perturbation resulting in the symmetry of the mean field for the composite network that depends on the sign of the interaction term.

The sign choice leading to anti-symmetry of the dynamic interaction between these two ATA DMM networks makes one think of the adolescent infighting between two political groups that are oppositely polarized. As long as one party holds a particular point of view the other party adopts the opposite point of view. Moreover there is no reconciling them, when a fluctuation induces one group to change its position this change immediately induces a corresponding reaction in the other group to change its position as well. Consequently, the weak coupling between the two groups prevents them from ever reaching accommodation.

Furthermore the interaction between two systems leads to an effective increase in the intra-cooperativity of each of them. The dynamics of coupled networks is reminiscent of the dynamics of a single ATA network with a cooperation parameter larger than the one characterizing the interacting subnetwork. Using the interpretation of political parties sharing the national stage, this corresponds to each of them becoming internally more uniform and orderly, eliminating diverse point of view within the group and becoming more radical in the opinions they hold.

In addition the anti-synchronization mechanism can also model desirable behavior. In the case of two well-behaved individuals carrying on a conversation, they politely take turns between talking and quietly listening. The distribution of time intervals between turn taking has been shown to be inverse power law \cite{pincus14} with an index that is larger than the value of 1.5 found in the coupled Langevin model. An index close to the value -2 that characterizes the optimum transfer of information between two complex networks as predicted by the PCM \cite{west08} was obtained by Abney \textit{et al}. \cite{abney14} for the inter-event distribution function. However detailed comparison of the present results to experiment remains to be systematically carried out.

\end{document}